\documentclass[journal]{IEEEtran}

\usepackage{amsmath}
\usepackage{amsfonts}
\usepackage{amsthm}

\usepackage{extarrows}

\usepackage{color}
\usepackage{tcolorbox}
\usepackage{colortbl}

\usepackage{longtable}
\usepackage{rotating}
\usepackage{multirow}
\usepackage{booktabs} 

\usepackage{makecell}

\usepackage{algorithmicx}
\usepackage{float}
\usepackage{algcompatible}

\usepackage[linesnumbered,ruled,vlined]{algorithm2e}
\SetAlCapNameFnt{\scriptsize}
\SetAlCapFnt{\scriptsize}

\SetAlgorithmName{Protocol}{Protocol}{List of Algorithms}

\tcbuselibrary{most}
\usepackage{mathrsfs}

\makeatletter
\newcommand{\thickhline}{%
    \noalign {\ifnum 0=`}\fi \hrule height 1pt
    \futurelet \reserved@a \@xhline
}
\newcolumntype{I}{!{\vrule width 1pt}}
\makeatother

\usepackage{cite}

\usepackage{comment}
\usepackage{amssymb,mathtools}
\usepackage{nccmath}
\usepackage{adjustbox}

\usepackage{graphicx}
\usepackage{subcaption}
\usepackage{caption}

\makeatletter
\renewcommand{\fnum@figure}{Fig. \thefigure}
\makeatother

\usepackage[english]{babel}
\usepackage[utf8]{inputenc}

\begin{document}

\author{Jianhua Li,~\IEEEmembership{Member,~IEEE,}
        Lingjuan Lyu,~\IEEEmembership{Member,~IEEE,}
        Ximeng Liu,~\IEEEmembership{Senior Member,~IEEE,}
        Xuyun Zhang,~\IEEEmembership{Member,~IEEE,}
        and~Xixiang~Lyu
\IEEEcompsocitemizethanks{\IEEEcompsocthanksitem This work was supported in part by the National Natural Science Foundation of China under Grant No. U1804263, No. 62072109, No. 62072356, and the Australia Research Council DECRA under Grant No. DE210101458. \protect\\

\IEEEcompsocthanksitem J. Li is with the UniCloud Australia, Melbourne, Australia, the State Key Laboratory of Integrated Services Networks, Xidian University, China, E-mail: jackleeml@hotmail.com. 
Lingjuan Lyu is with Ant Group, Hangzhou, China, Email: lingjuanlvsmile@gmail.com. 
Ximeng Liu is with College of MCS, Fuzhou University, the State Key Laboratory of Integrated Services Networks, Xidian University, China, Email: snbnix@gmail.com
Xuyun Zhang is with Department of Computing, Macquarie University, Australia, Email: xuyun.zhang@mq.edu.au. 
Xixiang Lyu, Xidian University, Xi'an, China, Email: xxlv@mail.xidian.edu.cn
}
}

\title{FLEAM: A Federated Learning Empowered Architecture to Mitigate DDoS in Industrial IoT}

\maketitle

\begin{abstract}
Due to resource constraints and working surroundings, many IIoT nodes are easily hacked and turn into zombies from which to launch attacks. It is challenging to detect such networked zombies rooted behind the Internet for any individual defender. We combine federated learning (FL) and fog/edge computing to combat malicious codes. Our protocol trains a global optimized model based on distributed datasets of collaborators while removing the data and communication constraints. The FL-based detection protocol maximizes the values of distributed data samples, resulting in an accurate model timely. On top of the protocol, we place mitigation intelligence in a distributed and collaborative manner. Our approach improves accuracy, eliminates mitigation time, and enlarges attackers' expense within a defense alliance. Comprehensive evaluations confirm that the cost incurred is 2.7 times larger, the mitigation response time is 72\% lower, and the accuracy is 47\% higher on average. Besides, the protocol evaluation shows the detection accuracy is approximately 98\% in the FL, which is almost the same as centralized training.
\end{abstract}

\begin{IEEEkeywords}
Federated Learning, Iterative model averaging (IMA), Gated recurrent unit GRU), Fog/edge, Industrial IoT DDoS, Cybersecurity.
\end{IEEEkeywords}

\IEEEpeerreviewmaketitle

\section{Introduction}

\IEEEPARstart{B}{ecause} of its potential to make better and faster decisions, the industrial Internet-of-things (IIoT) are reshaping many industries with connected sensors and machines. In modern transportation systems, engineers deploy the acoustic detection system to track bearing flaws in running trains and recognize dangerous conditions. To further expedite the decision-making process, fog/edge computing has become the game-changer for businesses to reduce latency and improve efficiency \cite{li2017virtual}. Compared with cloud servers, IIoT devices are easily hacked and turn into networked zombies, leading to notorious botnets and disasters \cite{kambourakis2017mirai}. Cybercriminals prefer to use such zombies to attack other entities for their stealth, large quantity, and low expense. Consequently, the distributed denial of service (DDoS) attacks have developed into a significant public hazard on the Internet  \cite{Ficco2015attack}.

Admittedly, machine learning (ML) improves the detection capacity of recognizing malicious traffics  \cite{Hussain2020detect} \cite{Abdollah2020detect} \cite{Farivar2020ai}. In general, ML relies on sufficient data samples to train a model with efficacy and accuracy. However, there are many concerns raised before the ML adoption in IIoT. First, many IIoT devices are resource-limited, almost incapable of running training tasks on the device. Also, many IIoT nodes are too application-specific to generate enough quantities or patterns of training samples \cite{akbar2018real}. IIoT threat landscapes are highly dynamic due to the surging growth of devices and applications every minute. A static well-trained model does not cope with quick change and unbalanced distribution. More interestingly, a centralized training model may become less accurate due to rapid changes in IIoT.

Motivated by the distribution of FL, fog, and edge computing, we argue that an effective combination of fog/edge computing and FL can defeat distributed attackers. Initialized by the ICT industry, fog/edge extends cloud computing to IIoT users and devices \cite{li2015ehopes}. FL is a decentralized learning process in which collaborators train a global model cooperatively \cite{lyu2020threats} \cite{uddin2021mutual}. Each participant downloads the global model from a server, retrains it locally with local data samples, and only sends model parameters to the cloud. Since malicious codes come from distributed bots, FL could update the model dynamically and detect attacking sources timely on fog nodes (FNs). Compared to centralized ML, FL has the benefits of lower latency, reduced traffic volumes, and close collaboration, making it more suitable for edge intelligence.

\begin{table}[t!]
\centering
\caption{Reported botnet-for-hire services}
\label{T1}
\begin{adjustbox}{max width=0.9\columnwidth}
\begin{tabular}{|l|l|c|c|}
\hline
\multicolumn{1}{|c|}{Botnet} & \multicolumn{1}{c|}{Bot Types} & \multicolumn{1}{c|}{Population} & \multicolumn{1}{c|}{Rental Price (\$)} \\ \hline
Botnet-Canada                & Computers                      & 1000                            & 270                                    \\ \hline
Botnet-the U. S.             & Computers                      & 1000                            & 180                                    \\ \hline
Botnet-the U. K.             & Computers                      & 1000                            & 240                                    \\ \hline
Botnet-France                & Computers                      & 1000                            & 200                                    \\ \hline
Boy Webcam                   & Hacked IIoT device              & 100                             & 1                                      \\ \hline
Girl Webcam                  & Hacked IIoT device              & 100                             & 100                                    \\ \hline
Remote controller               & Administration tool            & 1                            & 40                                     \\ \hline
\end{tabular}
\end{adjustbox}
\end{table}

Recently, DDoS attacks are becoming highly profitable with botnets. Botnet-for-hire services are available in many countries, as shown in Table I.\footnote{https://havocscope.com/black-market-prices/hackers/} Kaspersky experts estimated that the profit margin is up to 95\% above in one attack. Such profits drive numerous cybercriminals to keep attacking vulnerable IIoT nodes. Hence, we investigate the FL-empowered architecture to mitigate (FLEAM)  DDoS attacks. Our strategy is to make an attack more expensive than possible profits. In FLEAM, we place mitigation intelligence along the attacking path instead of deploying them tightly and heavily around a victim. Our contribution is fourfold:
\begin{itemize}
    \item We revisit DDoS attacks, pinpointing limitations of current solutions, such as insufficient consideration of attacker's expense, cross-defender collaborations, and mitigation strategy. Thus, we advocate joint defense and attacker-centric mitigation, preventing future attacks by incurring higher costs. 
    \item  An iterative model averaging (IMA) based gated recurrent unit (GRU) protocol is developed and testified to overcome diverse concerns raised in IIoT. The IMA-GRU protocol performs accurate detection on distributed data jointly.
    \item Grounded in the protocol, we place mitigation intelligence along attacking routes, incurring about 3.7 times expense as previously. 
    \item We showcase the benefits of FLEAM in contrast to classic solutions. The mitigation response time is about 72\% lower, while mitigation accuracy is 47\% higher approximately. Finally, we exhibit that our protocol has comparable accuracy to centralized training based on the UNSW NB15 dataset.
\end{itemize}

\begin{table}[]
\centering
\caption{Symbols and descriptions}
\label{T2}
\begin{adjustbox}{max width=0.9\columnwidth}
\begin{tabular}{|c|l|}
\hline
Symbols                                                                                   & \multicolumn{1}{c|}{Description}    \\ \hline
A                                                                                         & Attackable object, a possible victim during combat                   \\ \hline
AS                                                                                        & Analyzing submodule, a part of LAM performing local training                 \\ \hline
DS                                                                                        & Detection submodule, a part of LAM collecting data samples                 \\ \hline
DPM                                                                                       & DDoS policing module, the crux of local training and mitigation                \\ \hline
FS                                                                                        & Filtering submodule, a part of MPM with access control logic                 \\ \hline
IMA                                                                                       & Iterative model averaging           \\ \hline
LAM                                                                                       & Local analysis module               \\ \hline
MPM                                                                                       & Monitor and policing module      \\ \hline
MS                                                                                        & Monitoring submodule, a part of MPM monitoring and blocking data                \\ \hline
OF                                                                                        & Offensive firepower                 \\ \hline
TPD                                                                                       & Traffic policy database             \\ \hline
$\alpha_1, \alpha_2, \alpha_3, \alpha_4$ & Offensive power constants for both defenders and attacks             \\ \hline
$\Delta,   \delta_i$                                         & Baseline threshold                  \\ \hline
$\Gamma$        & Anomaly triggering threshold        \\ \hline
$\lambda$        & Regularizer controller        \\ \hline
$\tau$        & Mitigation time        \\ \hline
$\omega$        & Model parameter        \\ \hline
$\nabla \ell(\cdot;\cdot)$        & The gradient optimization function \\ \hline
$B$       & The batch size \\ \hline
$C, c_j$       & Malicious codes and malicious code $c_j$ \\ \hline
$D, D_i$       & The dataset and a local dataset \\ \hline
$E$            & Epoch \\ \hline
$F(\omega)$, $f(\omega)$        & General, local loss function        \\ \hline
$G$     & Global training model               \\ \hline
$h(\omega)$        & regularizer        \\ \hline
$I_{ij}$     & Idle resources $i$ on node $j$          \\ \hline
$M,m,i$        & Collaborators, number of collaborators, and their index                      \\ \hline
$H_i$        & The $i$-th node                      \\ \hline
$o_t$       & The GRU prediction \\ \hline
$P_i$        & Occurrence probability              \\ \hline
$Pkt_i$      & The $i$-th packet                     \\ \hline
$s, s_i$     & Packet symbol                       \\ \hline
$I_b$, $I_c$, $I_d$        & Betweenness centrality, closeness centrality, degree centrality               \\ \hline
$t$       & Training round                   \\ \hline
$TS, ts$    & Time window                         \\ \hline
$v_t$       & The GRU update function \\ \hline
$W, w_v$       & The IMA model weight, the GRU flow weight\\ \hline
$x$       & Feature space \\ \hline
\end{tabular}
\end{adjustbox}
\end{table}

The remainder of this paper is as follows. We present related work in Section II and the problem formulation in Section III. Section IV scrutinizes the FLEAM framework followed by the quantitative assessment in Section V. Section VI provides a detailed experimental evaluation of the FLEAM system and the IMA-GRU protocol. We finally conclude the paper in Section VII.

\section{Related Work}
Cybercriminals may try to overwhelm a victim or its surrounding infrastructure with botnets, causing service disruptions to legitimate users. Then, attackers demand payments for the ``protection'' of the victim. Due to the volume of attacks, a single defender finds it hard to defeat such joint attacks coordinated by the attacker. Attackers manage to keep back themselves as much as they can. Ficco \textit{et al.} showcased the benefits of stealthy attack patterns in \cite{Ficco2015attack}. The authors studied low-rate and intermittent flows and pinpointed the feature of the high stealthy and low cost of malicious traffic. The authors summarized that mitigation delay is critical for the success of protection. Similarly, Du \textit{et al.} explored how to identify honeypots in \cite{du2019honeypot} from an attacker's position. The authors showcased the interaction of attack and defense strategies.

ML has various benefits in analyzing DDoS traffics in a family of use cases. Hussain \textit{et al.} used deep learning to detect botnet traffic in cyber-physical systems \cite{Hussain2020detect}, while Abdollah \textit{et al.} applied ML to examine malicious traffic in Smart Grids \cite{Abdollah2020detect}. Farivar \textit{et al.} adopted AI to combat DDoS in IIoT \cite{Farivar2020ai}. In \cite{liu2019blockchain}, Liu \textit{et al.} investigated reinforcement learning to fight against DDOS. The above authors stress the feature of automation, low latency, high accuracy, and high scalability in dealing with attackers. 

Thanks to the programmability, global view, and centralized management, SDN could remove heavier reliance on other systems fighting against DDoS attacks \cite{wibowo2017multi}. Even so, SDN itself may introduce new DDoS attack surfaces like flow table overloading. Bhushan \textit{et al.} applied the queuing theory to keep minimal involvement of the SDN controller \cite{bhushan2019distributed}. Defenders may also adopt the reputation-based approach to limit traffic from malicious sources. Dahiya \textit{et al.} proposed a multi-attribute auction to mitigate DDoS in \cite{dahiya2020multi} \cite{dahiya2020qos}. Stergiou \textit{et al.} studied security and privacy issues in fog environments in \cite{stergiou2018security}, proposed adding a security `wall' between the cloud server and the Internet to offer a comprehensive security solution.

DDoS is detrimental to the interests of individuals, businesses, governments, and Internet service providers. Since an attacker coordinates distributed network zombies to attack a victim, defenders should collaborate to fight against DDoS efficiently. It requires defenders to share distributed botnets' information and actionable knowledge to beat common adversaries. Privacy does matter when personal data are being spied on, snooped, and mishandled. Existing solutions for privacy include adding noise to and using cryptosystem in data transferred between multiple parties. Such approaches often cause the loss of accuracy and cumulative delay. Grounded in federated learning (FL), our FLEAM only needs to upload model parameters to the cloud aggregator. Note model parameter sharing can reveal privacy, and we can use the existing solutions to protect privacy, but this is not our focus. Furthermore, in the DDoS detection context, the privacy issue is not a primary problem.

To this end, this work steps in, facilitating the joint detection and mitigation between all defenders. The victim has more accurate information, while the upstream defenders have a better position for mitigation placement. This research aims to detect and mitigate malicious codes at the proximity of the source end. The initial problem is to perform accurate and practical detection spanning multiple defenders. In reality, many IIoT devices suffer from fragile communication, insufficient computing power, and various constraints. It is less feasible to place training on them. Besides, fog gains its prosperity to expedite IIoT applications by distributing computing intelligence along the cloud-fog-device continuum \cite{li2017virtual}. We consider combining fog/edge with FL in the detection and mitigation due to their intrinsic characteristics, such as wide distribution and proximity to data sources. It is also more viable to place training intelligence on fog/edge as the fog has more data samples, better communication conditions, and its custodian role for IIoT \cite{li2021fast}.

\begin{figure}[!t]
\centering
\includegraphics[width=0.6\columnwidth]{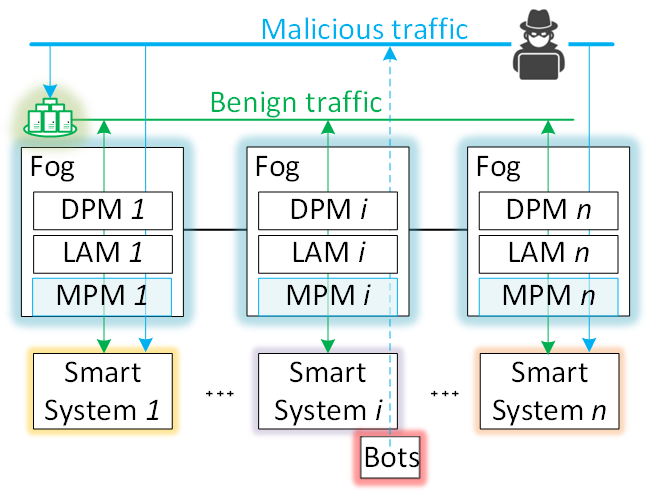}
\caption{The system model.}
\label{f1}
\end{figure}

\section{Problem Formulation}
In victim-centric solutions, detection and mitigation occur tightly around victims. Today, an attacker launches attacks from networked zombies rooted from other networks for stealth. Each defender performs detection and mitigation individually. Threats remain to other users until the bot is recognized and removed at the source end. We aim to remove such bots after it attacks a target in the defense alliance.

\subsection{System Model}
Fig. 1 displays our system model, where fog/edge connects IIoT systems. The monitor and policing module (MPM) focuses on monitoring and classifying inbound and outbound data. The local analysis module (LAM) monitors the health status of adjacent MPMs and responds to queries from MPM. Also, LAM performs traffic analysis under the coordination of the DDoS policy module (DPM). DPM incorporates a database of mitigation policy and system-specific training models. An attacker can hire botnets to attack any defender. Previously, a defender was only concerned about malicious codes targeting a local victim. The defender trains an individual model on the local dataset. Therefore, attackers can hide bots rooted as long as the bot does not attack a local target.

There are two training options for the alliance, including centralized training and FL on data from collaborators. Centralized training requires data sent to the cloud server, which triggers various issues such as high latency and resource limitations. Furthermore, IIoT threat landscapes are highly dynamic due to the surging growth of devices and applications every minute. More interestingly, some application-specific IIoT devices generate little traffic that might be insufficient to train a good model. Such participants can benefit from the FL for the optimal detection of threats. In FL, the cloud server sends the learning model to distributed participants. Each collaborator retrains the model based on local data. Training parameters are clustered and optimized on the cloud server, meanwhile removing the requirement of dataset uploading.

\subsection{Attack Model and Assumptions}
\subsubsection{Botnets} 
The botnet includes networked zombies rooted in distributed networks. An attacker can control bots to carry out passive and active attacks. The attacker launches DDoS to attack diverse victims like bandwidth and servers in the alliance.

\subsubsection{Defense Goals}
The first goal is to catch bot flows targeting a victim in the alliance. The second goal is to remove such bots rooted in a collaborator's network. 

\subsubsection{Offensive Firepower}
Cybercriminals attack a victim from distributed botnets, aiming to generate more output with the minimum input. We define the offensive firepower (OF) as: 
\begin{equation}
    OF_{bot} = \frac{Code_{out}}{Code_{in}} \gg 1 
\end{equation}
OF is generally inversely proportional to stealth. An attack needs to balance OF and stealth. We further assume fog/edge has the responsibility to manage and protect IIoT nodes.

\subsubsection{Secure Zone}
We assume that there are no malicious collaborators in the alliance. Thus, we can leverage FL to train an accurate model.

\subsubsection{Resource Allocation and consumption}
IIoT systems have many critical underlying resources like CPU and memory. Rapid depletion of such resources triggers system failure. During attacks, MPM notices the heavy consumption of some supplies. When a defender cannot allocate extra supplies to a victim, the victim is compromised.

\subsubsection{DDoS Concepts} We clarify several concepts in DDoS as follows.

\textbf{Active and passive mitigation}: A defender places detection and mitigation devices tightly around a victim, waiting for malicious codes to attack the victim. We name it the victim-centric model. In contrast, active mitigation refers to the distribution of mitigation and detection along attack routes, actively removing malicious codes. It is an attacker-centric model.

\textbf{Individual and joint defense}: In the individual defense model, a defender performs detection and mitigation on malicious activity imposed on local assets alone. Conversely, joint defense is a strategy for parties who share a common interest in cybersecurity against the common adversary. Collaborators swap defense information and actionable knowledge for DDoS mitigation.

\section{The FLEAM Architecture}
FLEAM empowers cross-defender detection and mitigation to strengthen overall security. Each collaborator needs to protect the objects within the alliance. 

\subsection{FLEAM Architecture Design}

There are four components of MPM, LAM, DPM, and the aggregator in FLEAM. In detail, MPM contains monitoring submodule (MS) and filtering submodule (FS). LAM includes detection submodule (DS) and analyzing submodule (AS). AS bears local training sends retrained parameters to the server. Traffic policy database (TPD) stores training models and results.

Fig. 2 presents the architecture details. MS rejects (flow 1) or forwards it (flow 2) to the FS in line with local pattern policy \cite{jia2016datapattern}. When no policy is available, MS forwards the data to DS (flow 3). DS categorizes the traffic using edge data analysis \cite{PRIYADARSHINI2019deeplearning}, responding to the query (flow 4). Meanwhile, DS notices AS of the suspicious flow (flow 5). AS uses the flow informed to retrain the global model and sends training results to DPM (flow 6). Next, TPD distributes policies to MS (flow 7) and FS (flow 8). Finally, FS rejects (flow 9) or sends (flow 10) the flow in line with the access control and routing logic. The logic is a sequence of local principles aligned with service-level agreements (SLAs). Finally, the trash bucket dumps everything in it.

\begin{figure}[!t]
\centering
\includegraphics[width=0.8\columnwidth]{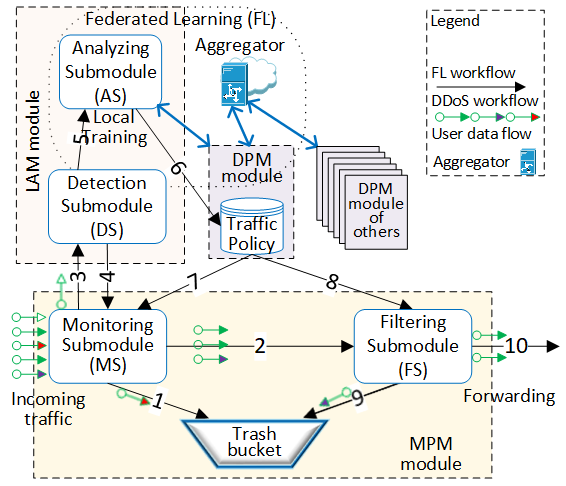}
\caption{The FLEAM architecture.}
\label{f2}
\end{figure}

Note DS makes a provisional decision to respond to MS, but it does not update policy. According to the policy, collaborators perform mitigation at local networks in near real-time. The actionable knowledge is the training result of Protocol 1.

\subsection{FL Design in FLEAM}
FLEAM includes three training elements, namely, AS, DPM, and aggregator. They are in a hierarchy of workers, coordinators, and the cloud server.

\subsubsection{System-specific Anomaly Detection}
Typically, a benign packet follows the regular pattern. We define a chain of data $Pkt_S$:
\begin{equation}
     Pkt_S = (Pkt_1, \cdots, Pkt_i, \cdots, Pkt_n) \;\;(i \subset n \in N_+)
\end{equation}

We profile incoming packets based on features in Table III below, where each row denotes a symbol.
\begin{equation*}
System_{prof} = 
\begin{pmatrix}
x_1^1 & x_2^1 & \cdots & x_9^1 \\
x_1^2 & x_2^2 & \cdots & x_9^2 \\
\vdots  & \vdots  & \ddots & \vdots  \\
x_1^n & x_2^n & \cdots & x_9^n
\end{pmatrix}
= \begin{pmatrix}
s_1\\
s_2\\
\vdots\\
s_n
\end{pmatrix}  
\end{equation*}
Then, we figure out the occurrence probability for each symbol using a gated recurrent unit (GRU) technique \cite{chung2014empirical}. As a particular variant of RNN, GRU has two gates, i.e., an input gate and a forget gate, making up an update gate in its structure. Compared to others, GRU performs well in less-frequent datasets.

\begin{table}[]
\centering
\caption{Packets and policy features}
\label{T3}
\begin{adjustbox}{max width=0.9\columnwidth}
\begin{tabular}{|l|l|l|}
\hline
\multicolumn{1}{|c|}{ID} & \multicolumn{1}{c|}{Feature} & \multicolumn{1}{c|}{Value}              \\ \hline
$x_1$                     & Application                  & The system coding features in IIoT       \\ \hline
$x_2$                     & Enterprise policy            & The enterprise policy in IIoT \\ \hline
$x_3$                     & Direction                    & Inbound or outbound                     \\ \hline
$x_4$                     & Destination IP and port      & IP and port number of destination       \\ \hline
$x_5$                     & Source IP                    & IP of the originator                     \\ \hline
$x_6$                     & Upper layer protocol         & Encapsulated protocol ID                \\ \hline
$x_7$                     & Packet length                 & Packet length indicator                     \\ \hline
$x_8$                     & Inter-arrival time           & Flow arrival time                       \\ \hline
$x_9$                     & Service-level agreement      & Fog platform SLA for IIoT              \\ \hline
\end{tabular}
\end{adjustbox}
\end{table}

\begin{equation}
    P_i = P(s_i|<s_{i-l},s_{i-l+1},\cdots,s_{i-1}>) \;\;(i>l \in N_+)
\end{equation}
where $P_i$ is the predicted possibility of $s_i$ for $Pkt_i$; $l$ is the length of preceding packets. The benign data usually follows a regular pattern. Before malicious traffic, we build a baseline $\Delta$ of occurrence possibility. It serves as the detection threshold. 
\begin{equation}
    \Delta = \{\delta_1, \cdots, \delta_i, \cdots, \delta_n\}  \;\;(i \subset n \in N_+)
\end{equation}

In sharp contrast to benign data, malicious traffic does not always follow the pattern, causing a smaller value of the occurrence possibility. We assume $Pkt_{j}$ is susceptible if $P_{j}$ meets 
\begin{equation}
    P_{j} < \delta_{j} \in \Delta  \;\;(j \in N_+)
\end{equation}
it reviews the packet is atypical. If AS receives $ts$ number of data packets in a timeslot $TS$, it classifies the flow into anomaly when:
\begin{equation}
     \begin{cases}
    TS = (Pkt_1, \cdots, Pkt_j, \cdots, Pkt_{ts}) \\
    \frac{s_j|P_j < \Delta}{ts} > \Gamma
    \end{cases}
      \;\;(j\subset ts \in N_+)
\end{equation}
where $\Gamma$ is the anomaly triggering threshold.

\subsubsection{Training Process}
In the joint detection framework, $m$ collaborators join the mitigation alliance using local data to perform stochastic gradient descent (SGD) optimization on the current model. The overall goal is to train the model across data samples of all collaborators $M$.

\begin{equation}
  \min_{\omega \in \mathbb{R}^d}F(\omega) = \sum_{i=1}^m \frac{\sum_{i \in D_i} f_i(\omega) + \lambda h(\omega)}{D} \,\, (\lambda \in [0,1]) 
  \end{equation}
where $F(\omega)$ is the general loss data function; $\omega \in \mathbb{R}^d$ is the local model parameter; $D_i$ is the size of the local dataset; $m$ is the number of collaborators; $f_i(\omega)$ is the local loss function; $h(\omega)$ is a regularizer, and $\lambda$ is a parameter that controls the importance of the regularizer.

The following equations present the functions of GRU used for training. 
\begin{equation}
    \begin{cases}
    v_t = \frac{w_v \cdot [o_{t-1},x_t] + x_t }{1+\exp(-t)} \\
    s_t = \frac{w_v \cdot [o_{t-1},x_t] + x_t }{1+\exp(-t)} \\    
    o'_t = \frac{(1-\exp(-2t)) \cdot W \cdot [s_t \odot o_{t-1}, x_t]}{1+\exp(2t)}\\
    o_t = (1-v_t) \odot o_{t-1} + v_t \odot o'_t 
    \end{cases}
\end{equation}
where $x$ is the feature space presented in Table III; $o_t$ is the prediction; $v_t$ is the update function; $w_v$ is the weight that manages the volume of information flowing through the GRU.

At the cloud server-side, the scheduler triggers a round of training. The aggregator invites collaborators through DPMs. The DPM makes sure that there is at least one local LAM is available. Otherwise, the DPM rejects the invitation. There are three steps involved as below when a DPM accepts the invitation.\\
\begin{enumerate}
    \item[Step1] The scheduler selects $m$ participants for training, and the collaborator sends the global model $G_t$ with parameter $\omega_0$ to such participants.
    \item[Step2] Each participant conducts retraining using data locally and updates $G_t$ for $E$ epochs of SGD with local batch size $B$ to obtain $\omega_{t+1}$.
    \item[Step3] Finally, the aggregator performs parameter convergence $G_{t+1}$ and broadcasts the new model to collaborators in an update process.
\end{enumerate}

Protocol 1 gives the details of the training process.

\begin{algorithm}[h]

\floatname{algorithm}{Protocol 1}
\caption{IMA-GRU protocol.}
\label{protocol1}
 \scriptsize
\begin{algorithmic}[1] 
 \STATEx {\textbf{Input}}: The $m$ collaborators are indexed by $i$; $B$ is the local batch size, $E$ is the number of epochs, $\theta$ is the learning rate, and $\nabla \ell(\cdot;\cdot)$ is the gradient optimization function. 
 \STATEx{\textbf{Output}}: Model parameter $\omega$.
\STATE{Server}: Initializes $\omega_0$, runs scheduler() and update() concurrently. 
\STATE{Scheduler}: Selects volunteers from the alliance through negotiations with DPM for this round of training. 
\STATE{DPM}: Checks the availability of the local LAM module. 
\IF{no LAM is available}
  \STATE reject the invitation
\ELSE
  \STATE confirm the willingness and select the LAM
\ENDIF
\STATE{Server}:  Gets the number of participants $i$
\FOR{ each round $t$ = $1,2,\cdots$}
  \STATE the server sends the global model and $\omega_0$ to participants
\ENDFOR   
\STATE{Updater}: Aggregates parameters received from the DPM of each participant 
\FOR{each participant in parallel} 
\STATE $\omega_{t+1} ^i \leftarrow$ ColUpdate($i,\omega_t$)
\STATE $\omega_{t+1} \leftarrow$ \,\, $\sum W_i \omega_{t+1}^i$
\ENDFOR
\STATE{Collaborator}: Facilitates the retraining
\IF{DPM has the model and $\omega_0$}
   \STATE distributes the job to the available LAM
\ENDIF
\STATE ColUpdate($i,\omega$): LAM retrain the model
\STATE The local data $D_i$ is split into batches of size $B$.
\FOR{each local epoch $i$ from 1 to $E$}
  \FORALL{batch $b \in$ $D_i$}
\STATE $\omega \leftarrow$ $\omega$ - $\theta \cdot \nabla \ell(\omega;b)$ 
  \ENDFOR
\STATE return $\omega$ to DPM and the server.
\ENDFOR
\end{algorithmic}
\end{algorithm}

During training, AS calculates the occurrence possibility and establishes the corresponding thresholds while the training result is the actionable policy. The outcome, together with the updated model, is stored in DPM.  After this, DPM updated parameters to the aggregator.

\subsubsection{The FL Implementation}
FNs bear AS and DPM while the cloud hosts the server. IIoT training data are generally non-IID, unbalanced, and massively distributed. We prefer to conduct the training at the fog/edge layer for two facts. One is that DDoS data against IIoT often come from clouds via fog/edge devices. Naturally, a fog node has more data samples than an IIoT device. The other is that fog/edge nodes have better communication conditions (fewer communication constraints) during the training. As such, we conduct the synchronous update scheme over selected participants round by round.

DPM adds flexibility to training as it coordinates multiple LAMs and only distributes training jobs to the appropriate LAM. Furthermore, DPM stores the latest training model and parameters, getting ready for new LAM modules for future training. Each LAM performs local computation based on its local dataset, preferred batch size, and the global state, updating DPM and the server with new parameters. Next, the server applies such updates to its optimal state.

The training result forms a DDoS policy stored in the traffic policy database that instructs MPM in joint detection and mitigation. Thanks to the separation of combat and training, an attacker cannot attack the training system directly. This mechanism limits the traffic entering the training gear but not affecting the training result because LAM has enough packet samples. We further demonstrate the training performance in Subsection VI.B.

We adopt the IMA-based synchronous approach as it has proven to be effective in the community \cite{mcmahan2017communication} \cite{zhao2018federated}. The SGD is a frequent optimizer, and the metric is the accuracy we used in our testbed. A local epoch is self-adaptive, according to the local dataset with the preferred batch size. The global training epoch is one aggregation.

\subsection{The Attacker-centric Mitigation}
Unlike victim-centric solutions where a defender passively waits for DDoS codes in the victim, attacker-centric solutions intend to place mitigation intelligence along the attacking path. Now, we investigate the placement strategy for active mitigation.
\subsubsection{Necessary Condition}
There must be at least one attackable entity in object aggregation $A$, namely, 
\begin{equation}
    A = \{A_1, A_2, \cdots, A_k, \cdots\} \;\;(A \not\in \emptyset, k \in N_+)
\end{equation}
Attackers carefully plan and select the valuable target(s) from $A$, for optimal return.
\begin{equation}
    Profit_{attack} = Value_{attack} - Cost_{attack}
\end{equation}
The second condition is the $Profit_{attack}$ must be positive.

The $Cost_{attack}$ may include finance, time, botnet expense, and others. To simplify our description, we only take into account the cost of botnets. Such payments cover rental and setup levy. 
\begin{equation}
    Cost_{bots} = Cost_{rental} + Cost_{setup}
\end{equation}

The least number per rental is one thousand bots. The ($Value_{attack}$) is the empirical reward for an attack. According to Equation (10), $Cost_{attack}$ is the determining factor. In DDoS, the winning party instigates the other party to spend more. 

Mitigation delay ($Time_{mtg}$) is critical for both parties, also the most significant factor in $Cost_{attack}$. Another one is the concrete population of bots ($N_b$) hired. The drop of mitigation delay results in the surge of $Cost_{attack}$.
\begin{equation}
    Cost_{attack} = \frac{1}{Kill_{pwr}} \cdot \frac{N_b}{Time_{mtg}} 
\end{equation}
where $Kill_{pwr}$ is the OF constant. The required number of $N_b$ is up to the supply of idle resource $I_{ij}$. 
\begin{equation}
    \begin{cases}
       \frac{dI_{ij}}{dt} &=  \alpha_1 I_{ij} - \alpha_2 I_{ij} N_b  \\
       \frac{dN_b}{dt} &=  \alpha_3 I_{ij} N_b - \alpha_4 N_b 
    \end{cases}
    \;\;(i,j \in N_+)  
\end{equation}
where $\frac{dI_{ij}}{dt}$ and $\frac{dN_b}{dt}$ describe the growth of populations; $t$ is time, $\alpha_1, \alpha_2, \alpha_3, \alpha_4$ are positive constants for killing powers. 

From Equation (12), we know mitigation delay is the most critical factor in DDoS.

\subsubsection{Node Selection}
Fog distributes required computing power along the cloud-fog-device continuum \cite{li2017virtual}. We can reduce the intensity of malicious traffic by placing more checkpoints along the attacking route. There are many heuristics to measure the degree of significance of fog nodes, including degree importance ($I_d$), betweenness importance ($I_b$), and closeness importance ($I_c$) \cite{laeuchli2021analysis}. In a series of network nodes $H$:
\begin{equation}
    H = \{ H_i\} \;\;(1<i \leq z \in N_+)
\end{equation}
where $z$ is the total number of nodes; $H_i$ is the $i$-th node in the network.
The $I_d$ counts the number of direct connections of each FN. 
\begin{equation}
    I_d = \frac{\sum_{i=1}^g[I_d(z^*)-I_d(i)]}{(z-1)(z-2)} \;\;(i,g,z \in N_+, z > 2)
\end{equation}
where $I_d(z^*)$ is the highest degree. The $I_d$ tells the level of difference in contrast with the highest degree.

The $I_b(i)$ measures the fragment of shortest paths via node $H_i$:
\begin{equation}
    I_b(i) = \sum_{j<k} \frac {H_{jk}(i)} {H_{jk}} \;\;(i,j,k \in N_+)
\end{equation}
where $H_{jk}$ is the number of shortest paths between nodes $H_j$ and $H_k$; $H_{jk}(i)$ is the count when $H_i$ is on the optimal route.

The $I_c(i)$ evaluates a node ($H_i$) per the average shortest interspace to other nodes:
\begin{equation}
    I_c(i) = \frac{1}{\sum_{j=1}^n d_{ij}} \;\;(i,j,z \in N_+)
\end{equation}
where $z$ is the number of FNs, and $d_{ij}$ is the shortest interspace from $H_i$ to $H_j$.

With the distribution of mitigation intelligence, FLEAM becomes attacker-centric at the fog/edge layer. Because attackers have little control over the attacking path, FLEAM can remove malicious data at the source end within the defender alliance. Next, we showcase its supreme performance compared with previous solutions.

\begin{table}[t!]
\caption{Various mitigation time for IIoT bots}
\centering
\label{T4}
\begin{adjustbox}{max width=0.9\columnwidth}
\begin{tabular}{|c|c|c|c|}
\hline
Bot Flows & Victim-centric (ms) & Attacker-centric (ms) & Interval (ms) \\ \hline
1         & 1900       & 600      & 5             \\ \hline
2         & 2800       & 800      & 10            \\ \hline
3         & 4400       & 1200     & 15            \\ \hline
\end{tabular}
\end{adjustbox}
\end{table}

\begin{figure}[!t]
\centering
\includegraphics[width=0.6\columnwidth]{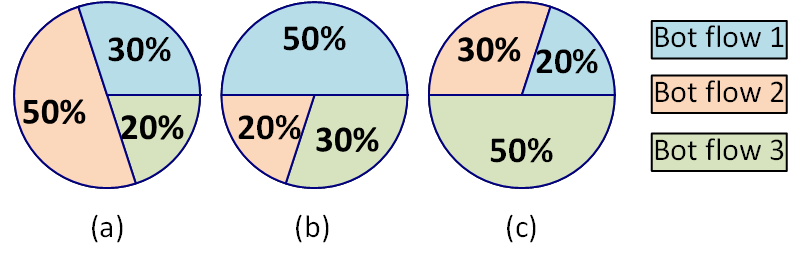}
\caption{Combined malicious traffic patterns, (a): Pattern-1, (b): Pattern-2, (c): Pattern-3.}
\label{f3}
\end{figure}

\section{Quantitative Evaluation}
During an attack, the attacker needs to balance the attacking power and stealth of DDoS codes. We simulate such malicious codes ($c_j$) by combining various patterns. Assuming Pattern-1 is with the most stealth and the least offensive strength while Pattern-3 is with the least stealth and the highest intensity. Bots of Pattern-2 are with medium power and fair stealth. Attackers try to outwit the detection intelligence with mixed patterns. A defender needs to spend $\tau_i$ time to defeat each $c_j$. The total mitigation delay $Time_{mtg}$: 
\begin{equation}
    Time_{mtg} = \sum_{i=0,j=1}^{m,3} \tau_i \cdot c_j  \;\;(i,j,m \in N_+)
\end{equation}

\begin{table}[t!]
\centering
\caption{The attacking expense in various countries}
\label{t5}
\begin{adjustbox}{max width=0.9\columnwidth}
\begin{tabular}{|l|c|c|c|c|}
\hline
\multicolumn{1}{|c|}{Botnets} & \begin{tabular}[c]{@{}c@{}}Classic model\\ (\$ per sec)\end{tabular} & \begin{tabular}[c]{@{}c@{}}FLEAM\\ (\$ per sec)\end{tabular} & \begin{tabular}[c]{@{}c@{}}Classic Model\\ (\$ per hour)\end{tabular} & \begin{tabular}[c]{@{}c@{}}FLEAM\\ (\$ per hour)\end{tabular} \\ \hline
Botnet-Canada                 & 0.157                                                                & 0.558                                                        & 0.566                                                                 & 2.009                                                         \\ \hline
Botnet-the U.K.               & 0.140                                                                & 0.496                                                        & 0.504                                                                 & 1.786                                                         \\ \hline
Botnet-France                 & 0.117                                                                & 0.413                                                        & 0.420                                                                 & 1.488                                                         \\ \hline
Botnet-the U.S.               & 0.105                                                                & 0.372                                                        & 0.378                                                                 & 1.340                                                         \\ \hline
\end{tabular}
\end{adjustbox}
\end{table}

\begin{figure}[t!]
\centering
  \begin{subfigure}{.25\textwidth}
  \centering
  \includegraphics[width=0.92\linewidth]{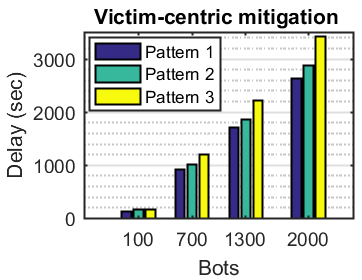}
  \caption{}
  \label{fig:sfig51}
\end{subfigure}%
\begin{subfigure}{.25\textwidth}
  \centering
  \includegraphics[width=0.92\linewidth]{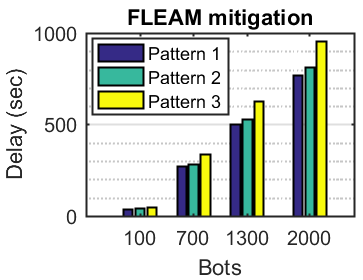}
  \caption{}
  \label{fig:sfig52}
\end{subfigure}%
\caption{Mitigation time in, (a): Victim-centric solutions, (b) FLEAM.}
\label{f4}
\end{figure}

We employ a Monte Carlo tool to witness the performance comparison between victim-centric solutions with FLEAM. Fig. 3 presents combined patterns of the above malicious codes. In more detail, bots generate DDoS traffic at random intervals of 5, 10, and 15 microseconds. It takes 7 microseconds to analyze each flow for a defender. We further adopt the average processing time shown in Table IV \cite{li2015ehopes}. 

Fig. 4 demonstrates the average mitigation delay as the bot population grows. Overall, the system response time gets longer with the increment of bots. Nevertheless, FLEAM reduces mitigation delay to approximately 28\% of which in the victim-centric model. Fig. 4(a) shows the mitigation response time is 1715.91 seconds per 1000 bots with victim-centric solutions. Fig. 4(b) illustrates the mitigation delay is 483.74 seconds in attacker-centric solutions (FLEAM). 

The rental price of 1000-bot botnets varies in different countries. Assuming the cybercriminal only attacks one victim, and the price is as in Table I. We compute the concrete expense per 1000-bot expense using Equation (12). The cost is about 0.1 dollars for Botnet-the US in the victim-centric model and 0.37 dollars per second in our FLEAM. FLEAM enlarges the expense to about 3.7 times.

It is of paramount importance for defenders to kill malicious codes as fast they can. Overall, FLEAM has the shortest mitigation delay. From an attacker's perspective, the defender reduces the survival time of bots significantly. Cybercriminals must invest more bots to overwhelming a victim in FLEAM. As such, the attacker will stop attacking the victim according to necessary conditions.

\section{Experimental Evaluation}
We now witness various levels of attacks and solutions to combat DDoS.
\subsection{System Evaluation}
\subsubsection{The Settings}
We set up the testbed with OPNET Modeler \cite{lu2012unlocking} to simulate the victim-centric and FLEAM mitigation process. An Ethernet workstation generates benign traffic from legitimate users. A 100BaseT Switched LAN simulates bot traffics. A firewall and a router operate as the mitigation and routing logic for a collaborator. The collaborator connects to a cloud with components of an IP cloud, a firewall, a server, and a reflector. FTP mimics benign traffic, while HTTP serves malicious codes. The three offensive power settings include Light, Medium, and Heavy.

\begin{figure}[t]
\centering
  \begin{subfigure}{.25\textwidth}
  \centering
  \includegraphics[width=0.92\linewidth]{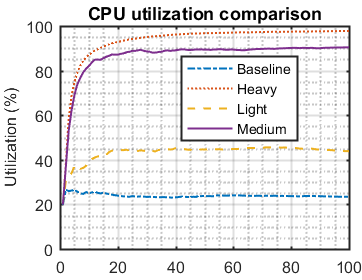}
  \caption{}
  \label{fig:sfig61}
  \end{subfigure}%
  \begin{subfigure}{.25\textwidth}
  \centering
  \includegraphics[width=0.92\linewidth]{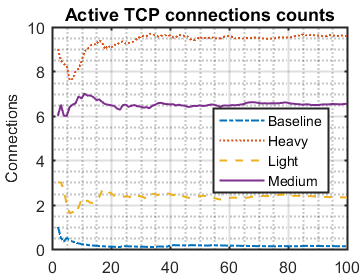}
  \caption{}
  \label{fig:sfig62}
  \end{subfigure}  
  \begin{subfigure}{.25\textwidth}
  \centering
  \includegraphics[width=0.92\linewidth]{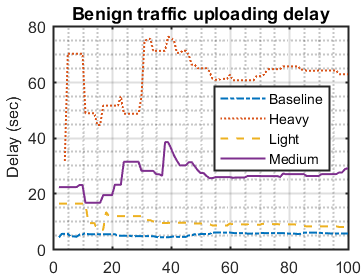}
  \caption{}
  \label{fig:sfig63}
  \end{subfigure}%
  \begin{subfigure}{.25\textwidth}
  \centering
  \includegraphics[width=0.92\linewidth]{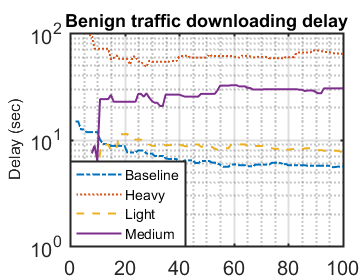}
  \caption{}
  \label{fig:sfig64}
  \end{subfigure}%
\caption{Various levels of attacks.}
\label{f5}
\end{figure}

Our experimental evaluation has six scenarios. In the first four scenarios, we observe the attacking process and levels, encompassing Baseline (no attack, benign traffic only), Light (malicious traffics under the threshold, the user experience is lightly affected), Medium (malicious data cause poor user experience), and Heavy (malicious traffic causes the severe failure of service provisioning). Fig. 5 (a) shows average CPU utilization of 20\%, 45\%, 85\%, and 98\% in the above scenarios. Fig. 5(b) illustrates active TCP connections triggered by botnets. Fig. 5(c) demonstrates the uploading delay for benign data. Fig. 5(d) displays the downloading delay. As can be seen, the attacker caused disconnections and service disruption via these attacks. 

Next, we start from the unprotected scenario (Heavy) to explore DDoS combat. The mitigation solutions include Unprotected (no protection), FLEAM (attacker-centric mitigation), and Classic (victim-centric mitigation). The fog firewall reproduces the FLEAM model. The cloud firewall replicates previous solutions.

\begin{figure}[t]
\centering
\begin{subfigure}{.25\textwidth}
  \centering
  \includegraphics[width=0.92\linewidth]{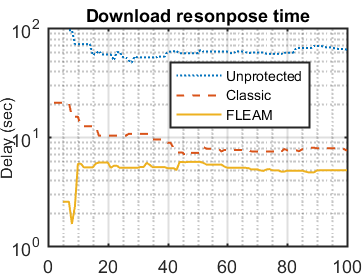}
  \caption{}
  \label{fig:sfig71}
\end{subfigure}%
\begin{subfigure}{.25\textwidth}
  \centering
  \includegraphics[width=0.92\linewidth]{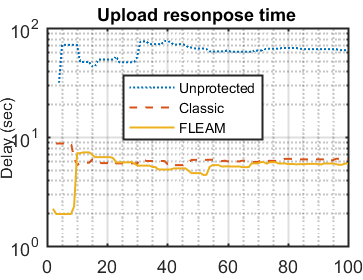}
  \caption{}
  \label{fig:sfig72}
\end{subfigure}
\begin{subfigure}{.25\textwidth}
  \centering
  \includegraphics[width=0.92\linewidth]{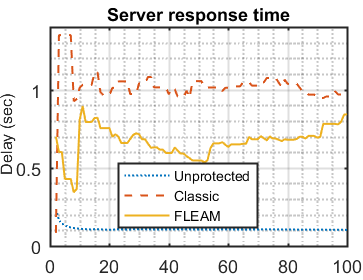}
  \caption{}
  \label{fig:sfig73}
\end{subfigure}%
\begin{subfigure}{.25\textwidth}
  \centering
  \includegraphics[width=0.92\linewidth]{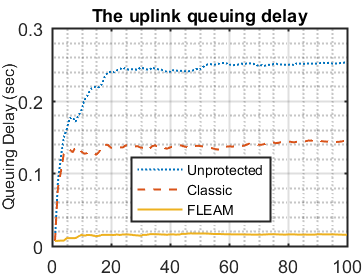}
  \caption{}
  \label{fig:sfig74}
\end{subfigure}
\begin{subfigure}{.25\textwidth}
  \centering
  \includegraphics[width=0.92\linewidth]{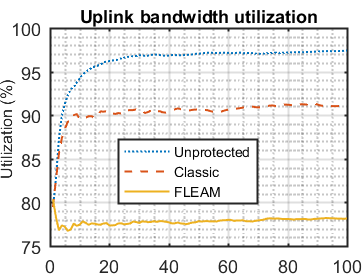}
  \caption{}
  \label{fig:sfig75}
\end{subfigure}%
\begin{subfigure}{.25\textwidth}
  \centering
  \includegraphics[width=0.92\linewidth]{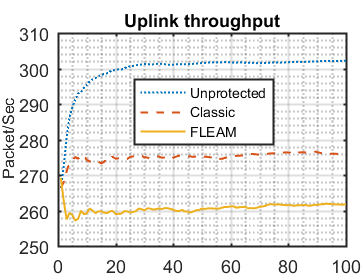}
  \caption{}
  \label{fig:sfig76}
\end{subfigure}
\caption{The performance comparison.}
\label{f6}
\end{figure}

\subsubsection{The Result Comparison}
The downloading delay, uploading delay, server response delay, uplink queuing time, bandwidth utilization, and throughput are collected and analyzed. Fig. 6(a) presents downloading delay in the above mitigation scenarios. Benign data delay is in a range of 50 and 100 seconds with the Unprotected case. A Classic solution reduces such delay to 6 to 10 seconds. In contrast, the benign delay is merely 0.7 to 5 seconds in FLEAM. Fig. 6(b) displays uploading delay where FLEAM performs slightly better than its counterpart. Following this, we measure the server TCP delay. Note the server TCP delay is about 0.1 seconds with the Unprotected, as revealed in Fig. 6(c). Protection causes the increment of server delay. Such delay is an acceptable cost due to reduced queuing time. Again, FLEAM performs better than previous solutions.

During combat, a client needs to cache outbound data in its queue. As shown in Fig. 6(d), the queuing delay is 0.2 seconds without protection. Legitimate users suffer from disconnection and interruptions, and the quality of experience is critically substandard. Benign traffic recovers when mitigation runs. We see the queuing delay in a range of 0.14 and 0.01 seconds with various protections. Fig. 6(e) shows the uploading bandwidth percentage. FLEAM saves bandwidth as it stops DDoS codes from entering the victim. Fig. 6(f) demonstrates the throughput of the bearing communication channel. We find there are about 260, 275, and 310 packets each second in FLEAM, Classic, and Unprotected cases.

After that, we measure system accuracy in benign data rate ($BPR$) and malicious data drop rate ($MDR$) \cite{alsirhani2018ddos}. Without cooperation, an individual solution cannot protect other users. FLEAM empowers cross-defender mitigation, leading to an attacker-centric solution. We harvest data sent, dropped packets, and packets received on all devices, getting $BPR$ and $MDR$. Then, we measure system accuracy as below.
\begin{equation}
    System_{accuracy} = \frac{BPR + MDR}{2}
\end{equation}

Fig. 7 demonstrates the  $System_{accuracy}$ in various settings. Fig. 7(a) tells individual mitigation accuracy is approximately 51\% and 52\% in the attacker-centric and victim-centric solutions. To compare with, Fig. 7(b) demonstrates the joint mitigation accuracy is much higher than an individual solution. The joint mitigation has up to 95\% accuracy in classic solutions and 98\% accuracy with FLEAM.

To sum up, FLEAM defeats DDoS threats accurately and quickly. It forces the attacker to invest more bots, discourages cost-sensitive cybercriminals from launching further attacks. 

\subsection{Protocol Evaluation}
Note the mitigation accuracy heavily relying on accurate detection. The outstanding performance of FLEAM depends on the detection empowered by our IMA-GRU protocol.

\subsubsection{The Protocol Testbed}
In this section, we test the protocol with the UNSW NB-15 dataset \cite{moustafa2015unsw}. It is a labeled data set for network intrusion detection systems with 49 features. The PySyft framework \cite{ryffel2018generic} is adopted to simulate four virtual workers, namely, four collaborators. PySyft library is a Python library for multiparty deep learning. We feed a random portion of the dataset to each collaborator, and then they start their training. The hidden layer of GRU is 100, and the epoch is 20. The testing dataset is 10\% of each local dataset randomly selected based on $k$-fold validation. The learning rate is 0.01. Depending on the volume of a dataset, the batch size varies from 32 to 1024. We conduct all experiments using Python 3.8.2 with Ubuntu 18.04.

\begin{figure}[t]
\centering
  \begin{subfigure}{.25\textwidth}
  \centering
  \includegraphics[width=0.92\linewidth]{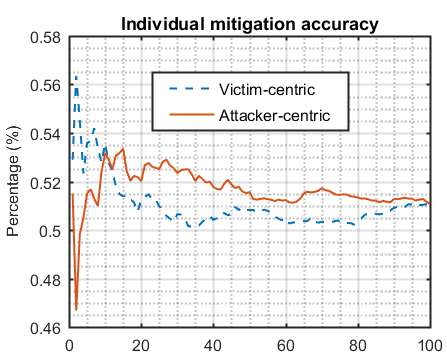}
  \caption{}
  \label{fig:sfig81}
  \end{subfigure}%
  \begin{subfigure}{.25\textwidth}
  \centering
  \includegraphics[width=0.92\linewidth]{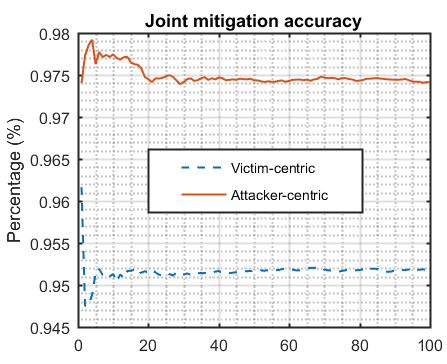}
  \caption{}
  \label{fig:sfig82}
  \end{subfigure}  
\caption{The system mitigation accuracy comparison.}
\label{f7}
\end{figure}

\begin{figure}[!t]
\centering
  \begin{subfigure}{.25\textwidth}
  \centering
  \includegraphics[width=0.92\linewidth]{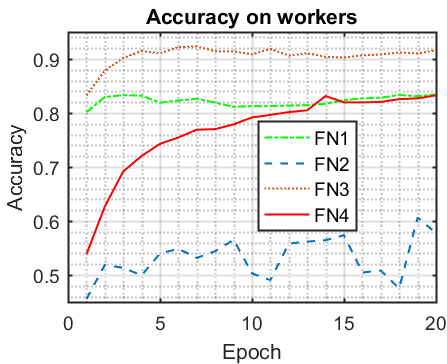}
  \caption{}
  \label{fig:sfig91}
  \end{subfigure}%
  \begin{subfigure}{.25\textwidth}
  \centering
  \includegraphics[width=0.92\linewidth]{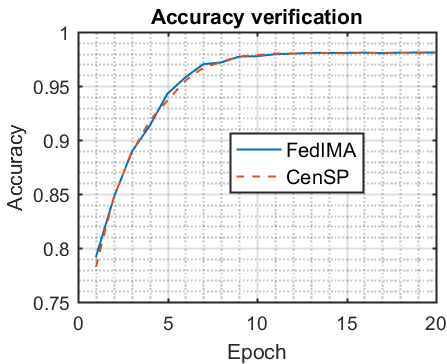}
  \caption{}
  \label{fig:sfig92}
  \end{subfigure}  
\caption{The protocol evaluation.}
\label{f8}
\end{figure}

\subsubsection{The Accuracy Comparison}
We first evaluate the accuracy of each collaborator when it conducts the individual training on the allocated dataset. Note each collaborator performs centralized training locally. Fig. 8(a) shows the result of each worker. In general, the accuracy varies significantly from one defender to another. For example, the accuracy on FN2 is oscillating around 52\% compared with 91\% on FN3. The reasons for various performances include an unbalanced dataset, training parameters, and capacity. Some IIoT users generate more data due to external and internal activities. Meanwhile, their budget on security may limit the concrete detection capacity of training gears. 

DDoS attacks come from networked zombies behind the Internet. It is almost impossible to collect all DDoS data and train a model in a centralized manner. Even if we do, we can only get a perfect model at that moment. In this experiment, we train our model on the entire UNSW NB-15 dataset for baseline. The FL training result is then compared with the baseline, as demonstrated in Fig. 8(b). The CentSP refers to the accuracy resulting from the centralized dataset. The FedIMA is of which from the FL. As can be seen, the difference is almost ignorable. 

However, the communication cost will be too high to achieve such centralized training in reality. It is more likely to conduct centralized training on a small portion of the entire dataset. Note each collaborator performs centralized training locally. Comparing Fig. 8(a) and 8(b), we find that FL does a better job as FL trains the model based on more data samples. Overall, the FL is more scalable and practical than centralized training for IIoT.

\section{Discussions and Conclusion}
\subsection{Discussion}
Nowadays, cybercriminals attack a victim from networked zombies rooted in distributed networks. Previous solutions focus on technology improvement while passively waiting for malicious codes at the victim end. An attacker coordinates bots behind the Internet while a defender fights against DDoS individually. Due to its resource constraints and enormous scale, attackers get a cheap army of IIoT botnets to attack a target at will, gaining outstanding profits from individual defenders. Fleam addresses such embarrassment by bringing attackers and defenders to the same surface within the defender alliance.  Table VI summarizes various characteristics of FLEAM and previous solutions.

Due to resource limitations and working conditions, machine learning is not directly applicable to IIoT devices. We investigate federated learning to train a global model for the alliance. The integration of fog/edge with federated learning maximizes the value of distributed datasets. FLEAM conducts training and protection at the fog/edge layer, tackling concerns raised by IIoT devices and data samples. Such edge AI eliminates communication costs, latency, and security issues. The rapid deployment of 5G will make FLEAM more feasible and favorable.  

The IIoT threat landscapes are highly dynamic. It is critically important to update the model with such evolving threats. FLEAM maintains accuracy with the IMA-GRU protocol. The protocol only needs collaborators to upload their retrained parameters to the server. The server will aggregate the uploaded parameters and update all collaborators with the latest model with limited parameter exchange. With the quick recognition of bots, collaborators kill the bot at their local network. Finally, botnets will be more expensive.  

One of the limitations of FL is to require frequent communication among nodes during the learning process. The fragile communication environment in IIoT can cause the loss of model updates and suboptimal global models. IIoT data sets are heterogeneous in size, distribution, and life cycle. We have mitigated this situation by placing the training intelligence on fog/edge devices. Thanks to their substantial computing power and communication intelligence, the training process can be smoothly achieved with more balanced data samples. In this work, we adopt the server-coordinator-worker structure, i.e., Aggregator-DPM-LAM hierarchies,  using DPM as a local training coordinator that supports multiple LAMs.

\subsection{Conclusion}
The DDoS game never ends on the Internet. FLEAM provides a sharper approach to combat DDoS in a cooperative style. Initially, the IMA-GRU protocol performs joint detection, achieving higher accuracy without moving training data to the cloud aggregator. The training intelligence locates in fog/edge nodes instead of IIoT devices, addressing concerns raised from IIoT data, computing power, and communication constraints. On top of the protocol, FLEAM enables the mitigation at the source end, reducing the mitigation delay and enlarging the attacking expense. The user experience is also improved.

\begin{table}[t!]
\centering
\caption{Characteristics comparison}
\label{T6}
\begin{adjustbox}{max width=0.9\columnwidth}
\begin{tabular}{|l|l|l|} 
\hline
Characteristics        & \multicolumn{1}{c|}{FLEAM}             & \multicolumn{1}{c|}{Classic}  \\ 
\hline
Detection Element       & Policy and packet symbols             & Packet patterns only          \\ 
\hline
Mitigation occurs           & Along attacking path             & At the victim end             \\ 
\hline
Solution Focus         & Technology and expense             & Technology only               \\ 
\hline
Intelligence Placement & Distributed, attacker-centric        & Centralized, victim-centric                   \\ 
\hline
Mitigation Strategy      & Collaborative and active               & Individual and passive                       \\ 
\hline
System Performance    & More accurate, faster & Less accurate, slow           \\
\hline
\end{tabular}
\end{adjustbox}
\end{table}

FLEAM is an attacker-centric model as it distributes mitigation intelligence along the attacking path to a victim. FLEAM pushes the front-line to the source, potentially removing zombies before attacking the second victim within the alliance. More importantly, FLEAM is a simple, scalable, and easily adoptable framework, able to create actionable intelligence at the source end.

We have investigated attack and mitigation strategy, detection protocol, and expense measurement for FLEAM. The quantitative assessment witnesses the advantage of FLEAM in mitigation delay and incurred expense. We have further conducted an experimental evaluation to prove our claims. The result shows an outstanding performance of FLEAM in front of previous solutions. The accuracy comparison testifies to the correctness of our protocol on the UNSW NB-15 dataset. Even though we limit our scope to DDoS, FLEAM sets the example in dealing with profit-driven and expense-sensitive attacks. We are going to implement FLEAM in the 5G environment in our future research.

\bibliographystyle{IEEEtran}
\bibliography{./ddos.bib}

\begin{IEEEbiography}[{\includegraphics[width=1in,height=1.25in,clip,keepaspectratio]{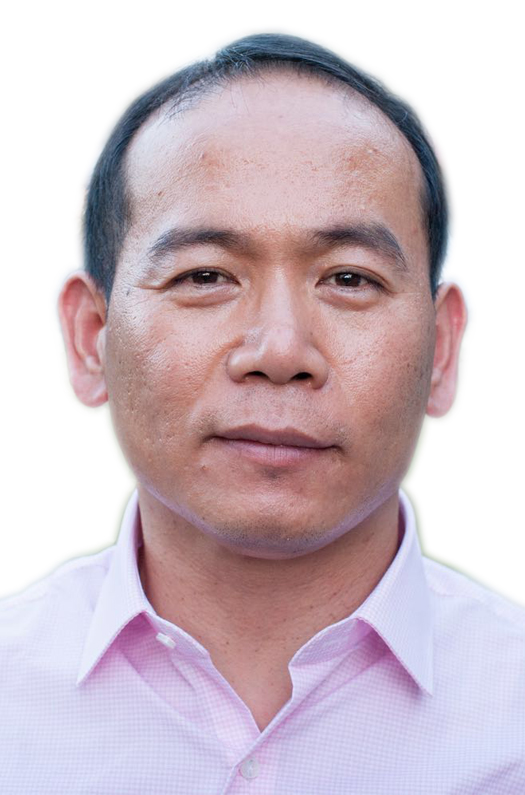}}]{Jianhua Li}
Jianhua Li received a Ph.D. degree in Information Engineering from the Swinburne University of Technology in 2019. He works for UniCloud Australia and Deakin University in Melbourne, Australia. He is also an ICT professional on multi-million-dollar projects and a global award winner in the ICT industry. His current research interests include cybersecurity, fog/edge computing, the Internet of things, cyber-physical systems, quality of experience, artificial intelligence, and networking softwarization. He has more than ten publications, including articles in IEEE Internet of Things Journal, IEEE Transactions on Industrial Informatics, and ELSEVIER Future Generation Computer Systems. Meanwhile, he is an active reviewer and TPC member for top-ranking journals and international conferences.
\end{IEEEbiography}

\begin{IEEEbiography}[{\includegraphics[width=1in,height=1.25in,clip,keepaspectratio]{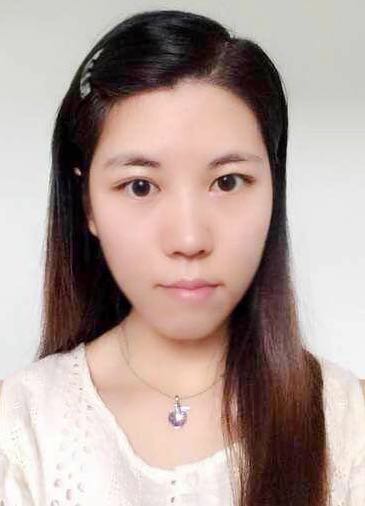}}]{Lingjuan Lyu}
Dr. Lingjuan Lyu is currently an expert researcher in Ant Group (Alipay). She was a Research Fellow with the National University of Singapore and a Research Fellow (Level B3, same level as lecturer/assistant professor) with the Australian National University. She received a Ph.D. degree from the University of Melbourne in 2018. Meanwhile, she was a winner of the IBM  Fellowship program (50 winners Worldwide) and contributed to various professional activities. Her current research interests span distributed machine/deep learning, privacy, robustness, fairness, and edge intelligence. She has publications in IJCAI, ICLR, SIGIR, ICASSP, EMNLP, NAACL, TII, JSAC, JIOT, TPDS, TDSC, etc. Her paper won the best paper award in FL-IJCAI’20.
\end{IEEEbiography}

\begin{IEEEbiography}[{\includegraphics[width=1in,height=1.25in,clip,keepaspectratio]{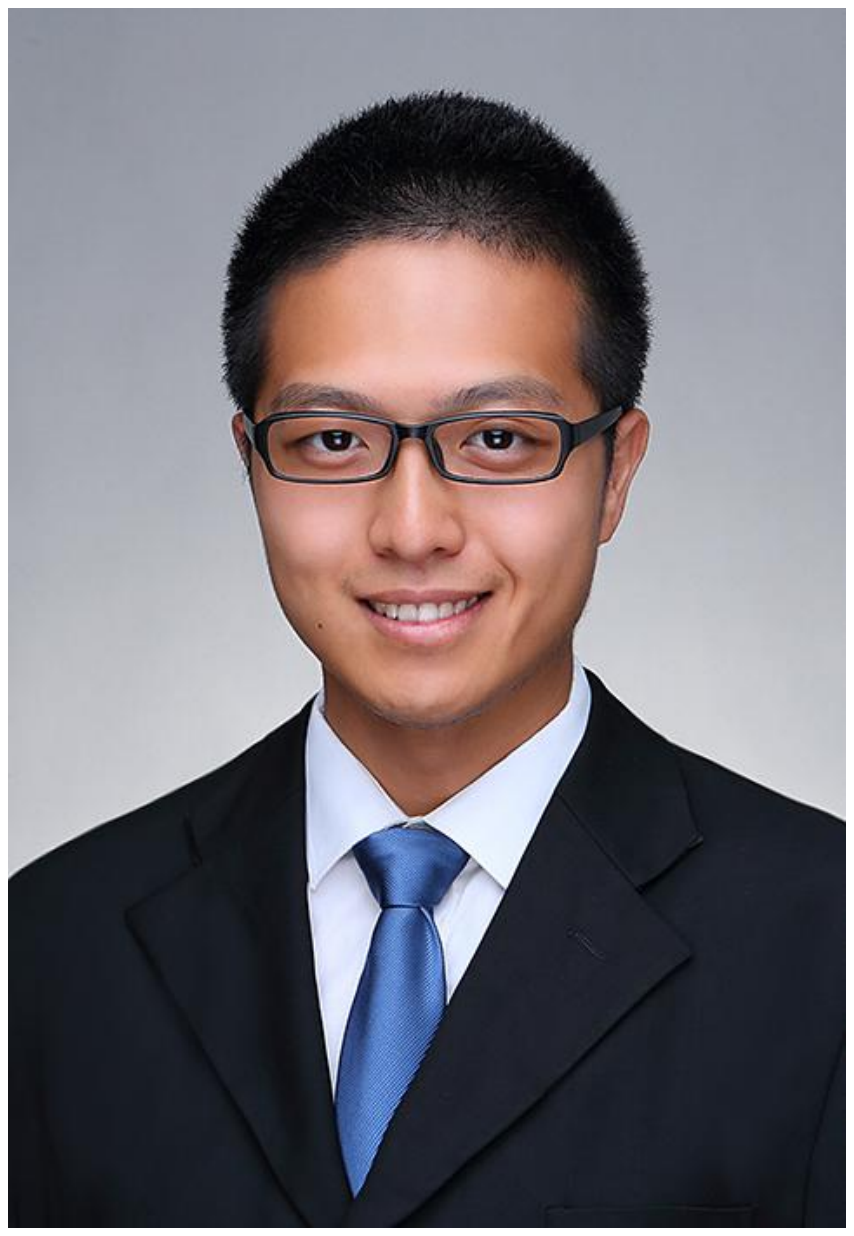}}]{Ximeng Liu}
Ximeng Liu (S' 13-M' 16-SM' 21) received the B.Sc. degree in electronic engineering from Xidian University, Xi’an, China, in 2010 and the Ph.D. degree in Cryptography from Xidian University, China, in 2015. Now he is the full professor in the College of Mathematics and Computer Science, Fuzhou University. He was a research fellow at the School of Information System, Singapore Management University, Singapore. He has published more than 250 papers on the topics of cloud security and big data security including
papers in IEEE TC, IEEE TIFS, IEEE TDSC, IEEE TPDS, IEEE TKDE, IEEE IoT Journal, and so on. He awards “Minjiang Scholars” Distinguished Professor, “Qishan Scholars” in Fuzhou University, and ACM SIGSAC China Rising Star Award (2018). His research interests include cloud security, applied cryptography and big data security. 
\end{IEEEbiography}

\begin{IEEEbiography}[{\includegraphics[width=1in,height=1.25in,clip,keepaspectratio]{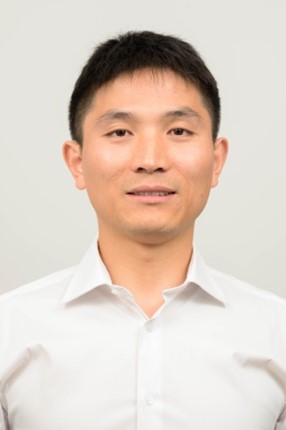}}]{Xuyun Zhang}
Dr Xuyun Zhang is currently working as a senior lecturer in Department of Computing at Macquarie University (Sydney, Australia). Besides, he has the working experience in University of Auckland and NICTA (now Data61, CSIRO). He received his Ph.D. degree in Computer and Information Science from University of Technology Sydney (UTS) in 2014, and his MEng and BSc degrees from Nanjing University. His research interests include scalable and secure machine learning, big data mining and analytics, cloud/edge/service computing and IoT, big data privacy and cyber security, etc. He is the recipient of 2021 ARC DECRA Award and several other prestigious awards. He has served as special issue guest editors for several high-quality journals like IEEE Trans. Industrial Informatics and IEEE Trans. Emerging Topics in Computational Intelligence.
\end{IEEEbiography}

\begin{IEEEbiography}[{\includegraphics[width=1in,height=1.25in,clip,keepaspectratio]{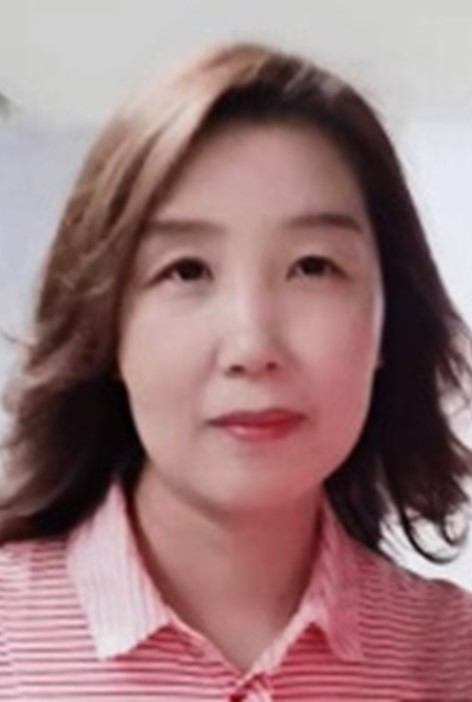}}]{Xixiang Lyu}
Xixiang Lyu studied at Xidian University from 1997 to 2007 and received her M.S. and Ph.D. degrees in 2004 and 2007, respectively. She is currently a full professor at Xidian University. Also, she is a member of the innovation team of the Ministry of Education on Key Technologies of Network and Information Security. Her research interests encompass anonymous networks, cryptographic algorithms, cybersecurity, privacy, and security in deep learning. In recent years, as the project leader, she has successively completed a project of National Key Basic Research Program of China, a Key Project of China National Natural Science Foundation, several general projects of China National Natural Science Foundation, and a project of Provincial Key R \& D projects, etc.
\end{IEEEbiography}

\end{document}